\documentclass[11pt]{article}
\usepackage[english]{babel}
\usepackage[latin1]{inputenc}
\usepackage{graphicx}
\usepackage{amsmath,amsfonts,amsthm}
\usepackage{color}

\title{\textbf{\Large Beta-binomial/gamma-Poisson regression models for repeated counts with random
parameters}}

\author{
     \large Mayra Ivanoff Lora \\
        \textit{Escola de Economia de São Paulo}\\
        \textit{Fundação Getulio Vargas, São Paulo, Brazil}\\[5pt]
     \large Julio M Singer \\
     \textit{Departamento de Estatística}\\
        \textit{Universidade de São Paulo, Brazil} \\
}
\date{}
\begin{document}
\maketitle

\begin{abstract}
\small {Beta-binomial/Poisson models have been used by many
authors to model multivariate count data.
Lora and Singer (Statistics in Medicine, 2008) extended such models to accommodate
repeated multivariate count data with overdipersion in the binomial component.
To overcome some of the limitations of that model, we consider a beta-binomial/gamma-Poisson alternative that also allows for both overdispersion and different covariances
between the Poisson counts.
We obtain maximum likelihood estimates for the parameters using a Newton-Raphson algorithm
and compare both models in a practical example.}
\end{abstract}


\section{Introduction}

Beta-binomial models have been used by many authors to model
binomial count data with different probabilities of success among
units from the same group of study. Williams (1975) used such
distributions to compare the number of fetal abnormalities of
pregnant rat females on a chemical diet during pregnancy to a
control group, both with fixed litter size. Gange et al. (1996)
analyzed the quality of health services (classified as appropriate
or not) during patient stay in a hospital using a similar
approach. To analyze mortality data in mouse litters with a fixed
number of implanted fetuses, Brooks et al. (1997) used such models
not only to allow for different probabilities of success among
units from the same group of study, but also to consider
overdispersion among them. Given that in many studies, the number
of trials may not be fixed, Comulada and Weiss (2007) considered a
multivariate Poisson distribution to model the number of successes
and failures in a random number of attempts, illustrating their
proposal with data from a HIV transmission study. Multivariate
Poisson distribution have also been used to model correlated count
data, as in Karlis and Ntzoufras (2003) who used such distribution
to model the number of goals of two competing teams.

In a study where the number of successes in a random number of
trials was observed repeatedly, and therefore are possibly
correlated, Lora and Singer (2008) consider multivariate
beta-binomial/Poisson models. In their proposal, the beta-binomial
component also accounts for overdispersion across units with the
same levels of covariates. The multivariate Poisson component
accommodates both the random number of trials and the repeated
measures nature of the data. The effect of possible covariates is
taken into account via the regression approach suggested by Ho and
Singer (1997, 2001). Their model, however, requires a constant
covariance term between the repeated number of trials and does not
allow for overdispersion in these counts. Since, as suggested by
Cox (1983), the precision of parameter estimates may be seriously
affected when overdispersion is not accounted for in the models
considered for analysis, we propose a beta-binomial/gamma-Poisson
model that not only incorporates such characteristics but is also
easier to implement computationally. The model, along with maximum
likelihood methods for estimation and testing purposes are
presented in Section 2. An illustration using data previously
analyzed by Lora and Singer (2008) is presented in Section 3. A
brief discussion and suggestions for future research are outlined
in Section 4.

\section{The beta-binomial/gamma-Poisson model for repeated measurements}

We denote the vector of responses for the $g$-th sample unit
($g=1,\dots,M$) by
\[
{\textbf Y_{g}}=(X_{g1},N_{g1},...,X_{gp},N_{gp})'
\]
with $X_{gh}$ corresponding to the number of successes in $N_{gh}$ trials performed under the $h$-th ($h=1,\dots,p$) observation condition. We assume that for all $g$ and $h$,
\begin{eqnarray}
&& X_{gh}\mid N_{gh}, \pi_{gh}  {\rm \; follow \; independent \;
binomial}
(N_{gh},\pi_{gh}) \; {\rm distributions} \\
&& \pi_{gh}
    {\rm \; follow \; independent \; Beta } (\mu(\textbf z_{\mu gh})/\theta(\textbf z_{\theta gh}),[1-\mu(\textbf z_{\mu gh})]/\theta(\textbf z_{\theta gh}))
     \; {\rm distributions} \\
&& N_{gh} \mid \tau_{g} {\rm \; follow \; independent \; Poisson}
(\lambda(\textbf z_{\lambda
gh}) \tau_{g}) \;{\rm distributions} \\
&& \tau_{g}  {\rm \; follow \; independent \; gamma}
(\alpha(\textbf z_{\alpha g})/\delta(\textbf z_{\delta
g}),1/\delta(\textbf z_{\delta g})) \; {\rm distributions}
\end{eqnarray}
where $\textbf z_{\mu gh}$, $\textbf z_{\theta gh}$, $\textbf z_{\lambda gh}$, $\textbf z_{\alpha g}$ and $\textbf z_{\delta g}$ are vectors of fixed covariates.

According to (1) and (2), the success probabilities may be different across units, but they are generated by beta distributions that may depend on covariates.
In (3) and (4), we follow Nelson (1985) to specify that the numbers of trials may also be different across units, but are generated by gamma distributions that may also depend on covariates.

The parametrizations ($0<\mu<1$, $\theta>0$) adopted in (2) and
($\alpha>0$, $\delta>0$) adopted in (4) are used to facilitate
maximum likelihood estimation, as suggested by Gange et al.
(1996); their relation to the usual beta($a, b$) parametrization,
as in Johnson and Kotz (1970), and the usual gamma$(c,d)$
parametrization, as in Mood et al. (1974), is given by
$$
\mu=\frac{a}{a+b} , \;\; \theta=\frac{1}{a+ b}, \;\;
\alpha=\frac{c}{d} \;\; {\rm and} \;\; \delta=\frac{1}{d}.
$$

The first and second order central moments of $\tau_{g}$ in (4)
are
\begin{eqnarray}
&&E(\tau_{g})=\alpha(\textbf z_{\alpha g}) \\
&&Var(\tau_{g})=\alpha(\textbf z_{\alpha g})\delta(\textbf
z_{\delta g})
\end{eqnarray}
From (3) and (4), the first and second order central moments of
the number of trials are
\begin{eqnarray}
&&E(N_{gh})=\lambda (\textbf z_{\lambda gh}) \alpha (\textbf
z_{\alpha g}) \\
&&Var(N_{gh})=\lambda (\textbf z_{\lambda gh}) \alpha (\textbf
z_{\alpha g}) \{1+\lambda (\textbf z_{\lambda gh}) \delta (\textbf
z_{\delta g}) \} \\
&& Cov(N_{gh},N_{gh'})=\lambda (\textbf z_{\lambda gh}) \lambda
(\textbf z_{\lambda gh'}) \alpha (\textbf z_{\alpha g}) \delta
(\textbf z_{\delta g})
\end{eqnarray}
for all $g,h,h' $, $h \neq h' $.
Similarly, the first and second order central moments of $\pi_{gh}$ in (2)
are
\begin{eqnarray}
&&E(\pi_{gh})=\mu(\textbf z_{\mu gh})\\
&&Var(\pi_{gh})=\mu(\textbf z_{\mu gh}) [1-\mu(\textbf z_{\mu
gh})]\theta(\textbf z_{\theta gh})[1+\theta(\textbf z_{\theta
gh})]^{-1}
\end{eqnarray}
Also, from (1) and (2), we may conclude that, for all $g$ and $h$,
\[
X_{gh} \mid N_{gh} \sim {\rm beta-binomial}[N_{gh},\mu(\textbf z_{\mu gh}),\theta(\textbf z_{\theta
gh})]
\]
with
\begin{eqnarray}
E(X_{gh})&=&\mu (\textbf z_{\mu gh}) \lambda (\textbf z_{\lambda
gh}) \alpha (\textbf z_{\alpha g}) \\ \nonumber Var(X_{gh})&=&\mu
(\textbf z_{\mu gh})[1-\mu (\textbf z_{\mu gh})]\frac{\theta
(\textbf z_{\theta gh})}{1+\theta (\textbf z_{\theta gh})}\lambda
^{2} (\textbf z_{\lambda gh})\alpha (\textbf z_{\alpha g})[\alpha
(\textbf z_{\alpha g})+\delta (\textbf z_{\delta g})]
\\ && + \mu (\textbf z_{\mu gh})\lambda (\textbf z_{\lambda gh})\alpha
(\textbf z_{\alpha g})[1+\mu (\textbf z_{\mu gh})\lambda (\textbf
z_{\lambda gh})\delta (\textbf z_{\delta g}) ] \\
Cov(X_{gh},X_{gh'})&=&\mu (\textbf z_{\mu gh}) \mu (\textbf z_{\mu
gh'}) \lambda (\textbf z_{\lambda gh}) \lambda (\textbf z_{\lambda
gh'}) \alpha (\textbf z_{\alpha g})  \delta (\textbf z_{\delta g})
\end{eqnarray}
for all $g,h,h'$, $h \neq h'$. The covariance between the numbers
of successes and trials is
\begin{eqnarray}
Cov(X_{gh},N_{gh})&=&\mu (\textbf z_{\mu gh}) \lambda (\textbf
z_{\lambda gh}) \alpha (\textbf z_{\alpha g}) \{1+ \lambda
(\textbf z_{\lambda gh}) \delta (\textbf z_{\delta g}) \}.
\end{eqnarray}

\noindent The parameters $\theta (\textbf z_{\theta gh})$ govern both the
variability of the success probabilities and the overdispersion of
the number of successes, that may also depend on the parameter
$\delta (\textbf z_{\delta g})$. When $\theta (\textbf z_{\theta
gh})$ and $\delta (\textbf z_{\delta g})$ are equal to zero, there
is no overdispersion for the number of successes. The parameters
$\delta (\textbf z_{\delta g})$ are also related to the
variability and overdispersion of the number of trials and to the
covariance between the numbers of trials and numbers of successes.
When $\delta (\textbf z_{\delta g})=0$, the repeated counts are
independent.

To investigate the effects of covariates, we adopt log-linear models of the form
\begin{eqnarray}
&&\mu(\textbf z_{\mu gh}) = \frac{\exp(\textbf z_{\mu gh}' \boldsymbol \beta_{\mu})}{1 + \exp(\textbf z_{\mu gh}' \boldsymbol \beta_{\mu})} \\
&&\theta(\textbf z_{\theta gh}) = \exp(\textbf z_{\theta gh}'\boldsymbol \beta_{\theta})\\
&&\lambda(\textbf z _{\lambda gh}) = \exp(\textbf z_{\lambda gh}'\boldsymbol \beta_{\lambda})\\
&&\alpha(\textbf z_{\alpha g}) = \exp(\textbf z_{\alpha g}'\boldsymbol \beta_{\alpha})\\
&&\delta(\textbf z_{\delta g}) = \exp(\textbf z_{\delta g
}'\boldsymbol \beta_{\delta})
\end{eqnarray}
where $\boldsymbol \beta_{\mu}$, $\boldsymbol \beta_{\theta}$,
$\boldsymbol \beta_{\lambda}$, $\boldsymbol \beta_{\alpha}$ and
$\boldsymbol \beta_{\delta}$ are vectors of parameters to be
estimated.

From (1), (2), (3) and (4) it follows that the joint probability mass
function for the number of trials and successes  for the $g$-th unit is
\begin{eqnarray} \nonumber
P(X_{g1},N_{g1},...,X_{gp},N_{gp})&=& \prod_{h=1}^{p} P(X_{gh}\mid
N_{gh}) P(N_{g1},...,N_{gp}) \
\\ \nonumber
\nonumber &=&\prod_{h=1}^{p} P(X_{gh}\mid N_{gh})
\left(\int_{0}^{\infty}\prod_{h=1}^{p}P(N_{gh}\mid
\tau_{g})f(\tau_{g})d\tau_{g}\right)
\end{eqnarray}
with $f$ denoting the density of (4). Since the logarithm of the
likelihood is given by
\begin{eqnarray}
\nonumber && logL(\boldsymbol \beta_{\mu},\boldsymbol
\beta_{\theta},\boldsymbol \beta_{\lambda},\boldsymbol
\beta_{\alpha},\boldsymbol \beta_{\delta})=
\\ \nonumber &=&\sum_{g=1}^{M}\sum_{h=1}^{p} log P(X_{gh} \mid
N_{gh},\boldsymbol \beta_{\mu},\boldsymbol \beta_{\theta}) +
\sum_{g=1}^{M} log P(N_{g1},...,N_{gp} \mid \boldsymbol
\beta_{\lambda},\boldsymbol \beta_{\alpha},\boldsymbol
\beta_{\delta}),
\end{eqnarray}
the parameters of the beta-binomial distribution ($\boldsymbol
\beta_{\mu}$,$\boldsymbol \beta_{\theta}$) can be estimated
separately from those of the gamma-Poisson distribution
($\boldsymbol \beta_{\lambda},\boldsymbol
\beta_{\alpha},\boldsymbol \beta_{\delta}$).

The beta-binomial probability mass function can be written as
\begin{eqnarray}
\nonumber && P(X_{gh}=x_{gh}\mid N_{gh}=n_{gh},\boldsymbol
\beta_{\mu},\boldsymbol \beta_{\theta})=
                         \begin{pmatrix}   n_{gh} \\   x_{gh} \end{pmatrix} \left\{\Gamma\left(\frac{1}{\theta(\textbf z_{\theta gh})}\right)
                          \left[\Gamma\left(\frac{1}{\theta(\textbf z_{\theta gh})} + n_{gh}\right)\right]^{-1}\right\} \\
            \nonumber &&\quad \times \left\{\Gamma\left(\frac{\mu(\textbf z_{\mu gh})}{\theta(\textbf z_{\theta gh})} + x_{gh}\right)
                                      \left[\Gamma\left(\frac{\mu(\textbf z_{\mu gh})}{\theta(\textbf z_{\theta gh})}\right)\right]^{-1}\right\} \\
            \nonumber && \quad \times \left\{\Gamma\left(\frac{1-\mu(\textbf z_{\mu gh})}{\theta(\textbf z_{\theta gh})} + n_{gh} - x_{gh}\right)
                                       \left[\Gamma\left(\frac{1-\mu(\textbf z_{\mu gh})}{\theta(\textbf z_{\theta gh})}\right)\right]^{-1}\right\}
                                          \\
            \nonumber &&=  \begin{pmatrix} n_{gh} \\  x_{gh} \end{pmatrix} \prod_{u=0}^{n_{gh}-1}[1 + u\theta(\textbf z_{\theta gh})]^{-1}\prod_{v=0}^{x_{gh}-1}[\mu(\textbf z_{\mu gh})
                                                     + v \theta(\textbf z_{\theta gh})] \\
                      &&\quad \times \prod_{w=0}^{n_{gh} - x_{gh}-1}[1-\mu(\textbf z_{\mu gh}) + w\theta(\textbf z_{\theta gh})]
\end{eqnarray}
where $\Gamma(r) = \int_{0}^{\infty} t^{r-1} e^{-t} dt$. The
expressions involving ratios between two gamma functions
(presented within brackets) make sense when $n_{gh} \neq 0$ (in
the first ratio), $x_{gh} \neq 0$ (in the second ratio) and
$x_{gh} \neq n_{gh}$ (in the third ratio). When these conditions
are not satisfied, the ratios between the gamma functions may be set
equal to one, and do not affect the conditional probability of $X_{gh}$ given
$N_{gh},\boldsymbol \beta_{\mu},\boldsymbol \beta_{\theta}$.

The kernel of the beta-binomial log-likelihood function is
\begin{eqnarray}
\nonumber
&L(\boldsymbol \beta_{\mu}, \boldsymbol \beta_{\theta})=&
\sum_{g=1}^{M}\sum_{h=1}^{p}\left[ \sum_{v=0}^{x_{gh}-1}log[\mu(\textbf z_{\mu gh})+v \theta(\textbf z_{\theta gh})]+ \right. \\
&&\left. \sum_{w=0}^{n_{gh}-x_{gh}-1} log[1-\mu(\textbf z_{\mu
gh})+w\theta(\textbf z_{\theta gh})]-
        \sum_{u=0}^{n_{gh}-1}log[1+u\theta(\textbf z_{\theta gh})]
       \right]
\end{eqnarray}

\noindent and we may use maximum likelihood methods adopting a
Newton-Raphson iterative process to estimate $\boldsymbol
\beta_{\mu}$ and $\boldsymbol \beta_{\theta}.$ The first and
second derivatives of (22) are shown in Lora and Singer (2008).
Method of moments estimates based on the beta-binomial
distribution may be used as initial values for $\mu(\textbf z_{\mu
gh})$ and $\theta(\textbf z_{\theta gh})$, as suggested by
Griffiths (1973). Likelihood ratio tests may be employed for model
reduction purposes, i.e., for constructing a parsimonious model
that captures the explainable variability in the data. For
example, to verify if the $q$-parameter vector $\boldsymbol
\beta^{*}$ is null, the test statistics $LR = 2(L-L^{*})$, with
$L^{*}$ indicating the log-likelihood under $H_{0}$ and $L$, this
logarithm under the alternative hypothesis may be employed.
Asymptotically, $LR$ follows a chi-squared distribution with $q$
degrees of freedom under the null hypothesis.

The probability function for the repeated number of trials based in (3) and (4) is
\begin{eqnarray} \nonumber
&& P(N_{g1}=n_{g1},...,N_{gp}=n_{gp}|\boldsymbol \beta_{\lambda},
\boldsymbol \beta_{\alpha}, \boldsymbol \beta_{\delta})=
\\ \nonumber
&& =\prod_{h=1}^{p} \left \{ \frac{[\lambda (\textbf z_{\lambda
gh})]^{n_{gh}}}{n_{gh}!} \right \} \left[\frac{1}{\delta (\textbf
z_{\delta g})}\right]^{\alpha (\textbf z_{\alpha g})/\delta
(\textbf z_{\delta g})}\Gamma\left(\sum_{h=1}^{p}
n_{gh}+\frac{\alpha (\textbf z_{\alpha g})}{\delta (\textbf
z_{\delta g})}\right) \left \{ \Gamma \left( \frac{\alpha (\textbf
z_{\alpha g})}{\delta (\textbf z_{\delta g})} \right) \right
\}^{-1}
\\ \nonumber
&& \div \left[ \sum_{h=1}^{p} \lambda (\textbf z_{\lambda gh})+
\frac{1} {\delta (\textbf z_{\delta g})} \right]^{\Sigma_{h=1}^{p}
n_{gh}+\alpha (\textbf z_{\alpha g})/ \delta (\textbf z_{\delta
g})}
\\ \nonumber && = \prod_{h=1}^{p} \left \{ \frac{[\lambda (\textbf
z_{\lambda gh})]^{n_{gh}}}{n_{gh}!} \right \}
\prod_{u=0}^{\Sigma_{h=1}^{p}n_{gh}-1} \left [ \alpha (\textbf
z_{\alpha
g})+u \delta (\textbf z_{\delta g}) \right ] \\
&& \div \left \{ \delta (\textbf z_{\delta g}) \left[
\sum_{h=1}^{p} \lambda (\textbf z_{\lambda gh}) \right] + 1 \right
\} ^{\Sigma_{h=1}^{p} n_{gh} +\alpha (\textbf z_{\alpha g})/
\delta (\textbf z_{\delta g})}
\end{eqnarray}

\noindent In (23), the simplifications for the rations between two gamma
functions make sense when $\sum_{h=1}^{p} n_{gh}\neq 0$. When this
condition is not satisfied, the ratio is also set equal to one, and it
does not affect the probability value.

The kernel of the gamma-Poisson log-likelihood function is
\begin{eqnarray} \nonumber
&&L(\boldsymbol \beta_{\lambda},\boldsymbol
\beta_{\alpha},\boldsymbol \beta_{\delta})= \sum_{g=1}^{M} \left
\{ \sum_{h=1}^{p} [n_{gh} log \lambda (\textbf z_{\lambda gh})]  +
\sum_{u=0}^{\Sigma_{h=1}^{p}n_{gh}-1}
log[\alpha (\textbf z_{\alpha g})+u \delta (\textbf z_{\delta g})] \right. \\
&&\left. - \left [\sum_{h=1}^{p} n_{gh} + \frac{\alpha (\textbf
z_{\alpha g})}{\delta (\textbf z_{\delta g})} \right] log
\left[\delta (\textbf z_{\delta g}) \left (\sum_{h=1}^{p} \lambda
(\textbf z_{\lambda gh}) \right)+1 \right] \right \}
\end{eqnarray}

\noindent and we adopt the same methods used with the
beta-binomial model to estimate $\boldsymbol
\beta_{\lambda},\boldsymbol \beta_{\alpha}$ and $\boldsymbol
\beta_{\delta}.$ The first and second derivatives of (24) are
shown at the Appendix. Method of moments estimates may be used
used as the initial values for $\lambda_{gh}(\textbf
z_{\lambda})$, $\alpha_{g}(\textbf z_{\alpha})$ and
$\delta(\textbf z_{\delta})$ here, too. Likelihood ratio tests may
be employed for model reduction purpose, along similar lines as
those considered for the beta-binomial model.

Both iterative processes are implemented in the R software and
the corresponding code can be downloaded from
http://www.ime.usp.br/$\sim$jmsinger.

\section{Data analysis}

To compare the beta-binomial/gamma-Poisson to the multivariate
beta-binomial/Poisson model, we consider the same data presented
in Lora and Singer (2008) from a study conducted at the Learning
Laboratory of the Department of Physiotherapy, Phonotherapy and
Occupational Therapy of the University of São Paulo, Brazil, to
evaluate the performance of some motor activities of Parkinson's
disease patients. For the sake of completeness, we repeat the
description of the study here. Twenty five patients with confirmed
clinical diagnosis of Parkinson's disease and twenty one normal
(without any preceding neurologic alterations) subjects repeated
two sequences of specified opposed finger movements (touching one
of the other four fingers with the thumb) during one minute
periods, with both hands. This was done both before and after a
four-week experimental period in which only one of the sequences
was trained (active sequence) with one of the hands; the other
sequence was not trained (control sequence). Half of the subjects
in each group trained the preferred hand (right for the
right-handed and left for the left-handed in the normal group or
the less affected by the disease in the experimental group) and
the other half trained the non-preferred hand. Information on the
number of attempted and successful trials were recorded with a
special device attached to a computer.

Six subgroups may be characterized by the combination of
disease stage (normal, initial or advanced) and use of the
preferred hand (yes or no). The repeated measures are
characterized by the cross-classification of the levels of
sequence (control or active) and evaluation session (baseline or
final). The specific objective of the study was to evaluate
whether training is associated with increases in the expected number
of attempted trials per minute (agility) and/or on the probability
of successful trials (ability). Note that the treatment could
improve agility without improving ability, so an evaluation of its
effect on both characteristics is important.

The means and variances of the number of attempted and successful
trials at the baseline and final evaluations with the active and
control sequences for patients at the different disease stages
using the preferred or non-preferred hands are presented in Table
1. Variances, instead of standard deviations, are displayed to
facilitate identification of overdispersion in the sense referred
by Nelder and McCullagh (1989), i.e., cases where variances are
greater than expected under Poisson or binomial distributions.
Overdispersion in the number of attempts, under a Poisson
distribution is clearly identified by comparing the observed mean
and variance; for the number of successes, on the other hand, it
is necessary to compare the observed and expected variances under
the binomial distribution ($np(1-p)$). For example, considering
normal subjects performing the active sequence at the baseline
session using the preferred hand, the expected variance under the
binomial model is $1.4$, while the observed variance is $49.0$,
highlighting the overdispersion for these counts too.

Correlation coefficients for the within-subject responses for the
normal patients using the preferred hand are displayed in Table 2.
For this subgroup, only $3$ out of the $28$ observed correlations
are smaller than $0.60$; this suggests that the counts are
probably related and it is sensible to use a model that can
accommodate this relationship. The correlation patterns for the
other subgroups are similar and are not presented.

\begin{table}[p]
\caption{Mean and variance (within parentheses) of the number of
attempted and successful trials.} \small \centering
\begin{tabular}{cccccc} \hline
  Disease & Evaluation & Intervention & Sequence & Successes & Attempts  \\
  stage   & session    & hand         &          &           &         \\\hline
  Normal  & Baseline & Preferred & Control & 17.1 (49.0) & 18.6 (46,2) \\
  Normal  & Baseline & Preferred & Active & 17.1 (72.3) & 17.9 (79.2) \\
  Normal  & Baseline & Non-preferred & Control & 18.1 (27.0) & 20.9 (47.6) \\
  Normal  & Baseline & Non-preferred & Active & 17.1 (37.2) & 19.5 (53.3) \\
  Normal  & Final & Preferred & Control & 20.9 (90.3) & 26.1 (44.9) \\
  Normal  & Final & Preferred & Active & 32.7 (139.2) & 33.1 (132.3) \\
  Normal  & Final & Non-preferred & Control & 24.2 (25.0) & 28.6 (38.4) \\
  Normal  & Final & Non-preferred & Active & 32.8 (74.0) & 34.4 (72.3) \\ \hline
  Initial & Baseline & Preferred & Control & 13.7 (24.0) & 16.3 (44.9) \\
  Initial & Baseline & Preferred & Active & 12.0 (23.0) & 13.5 (23.0) \\
  Initial & Baseline & Non-preferred & Control & 12.0 (17.6) & 14.6 (9.0) \\
  Initial & Baseline & Non-preferred & Active & 10.7 (20.3) & 13.6 (10.9) \\
  Initial & Final & Preferred & Control & 13.2 (30.3) & 16.8 (43.6) \\
  Initial & Final & Preferred & Active & 20.2 (9.6) & 21.8 (2.9) \\
  Initial & Final & Non-preferred & Control & 15.3 (112.4) & 20.3 (116.6)\\
  Initial & Final & Non-preferred & Active & 20.1 (33.6) & 20.4 (39.7) \\\hline
  Advanced & Baseline & Preferred & Control & 4.8 (22.1) & 7.1 (11.6) \\
  Advanced & Baseline & Preferred & Active & 4.6 (11.6) & 7.9 (14.4) \\
  Advanced & Baseline & Non-preferred & Control & 8.3 (72.3) & 12.5 (15.2) \\
  Advanced & Baseline & Non-preferred & Active & 13.5 (92.2) & 15.5 (57.8) \\
  Advanced & Final & Preferred & Control & 7.4 (75.7) & 11.9 (67.2) \\
  Advanced & Final & Preferred & Active & 13.5 (90.3) & 14.9 (77.4) \\
  Advanced & Final & Non-preferred & Control & 5.8 (31.4) & 12.8 (12.3) \\
  Advanced & Final & Non-preferred & Active & 22.5 (75.7) & 23.8 (75.7) \\ \hline
\end{tabular}
\end{table}

\begin{table}[h]
\caption{Correlation coefficients for the within-subject responses
for the normal subjects using the preferred hand} \vspace{0.3cm}
\small \centering
\begin{tabular}{cccccccccccc} \hline
   & & & \multicolumn{4}{c}{Baseline session} &\quad \qquad& \multicolumn{4}{c}{Final session} \\
   & & & \multicolumn{2}{c}{Active seq.}& \multicolumn{2}{c}{Control
   seq.}&&   \multicolumn{2}{c}{Active seq.}& \multicolumn{2}{c}{Control seq.}  \\
   & & & Suc. & Att. & Suc. & Att. & &Suc. & Att. & Suc. & Att.  \\ \hline
   Baseline & Active & Suc. & 1 & & & &&  &  & &  \\
   session & seq. & Att. & 0.99 & 1& & &&  &  & &  \\
   & Control & Suc. & 0.85 & 0.84 & 1& &&  &  & &  \\
   & seq. & Att. & 0.78 & 0.80 & 0.96 & 1&&  &  & &  \\
   &&&&&&&&&& \\
   Final & Active & Suc. & 0.76 & 0.76 & 0.61 & 0.61&& 1 &  & &  \\
   session & seq. & Att. & 0.74 & 0.74 & 0.61 & 0.63& & 0.99&  1& &  \\
   & Control & Suc. & 0.53 & 0.49 & 0.59&0.63 & &0.60& 0.61 & 1 &   \\
   & seq. & Att. & 0.81 & 0.82 & 0.70& 0.69& & 0.93& 0.92 & 0.50& 1 \\ \hline
\end{tabular}
\begin{center}
Codes: Suc.$=$Successes, Att.$=$Attempts and seq.$=$sequence
\end{center}
\end{table}

The analysis strategy consisted in fitting initial models of the
form (16)-(20) with all main effects and first order interactions,
and trying to reduce them by sequentially eliminating the
non-significant terms. The parameters are indexed by disease stage
(0=normal, 1=initial, 2=advanced), intervention hand (P=preferred,
N=non-preferred), evaluation session (B=baseline, F=final) and
sequence (C=control, A=active). We adopted a reference cell
parameterization with the reference cell corresponding to the normal
group (0), performing the active sequence (A) with the preferred
hand (P) at the baseline evaluation (B).

\subsection{Modelling the expected probability and dispersion of successful attempts}

For both beta-binomial/gamma-Poisson and multivariate beta-binomial/Poisson
models, the parameters of the beta-binomial components can
be estimated separately from those of the gamma-Poisson or the multivariate
Poisson distributions. Therefore, modelling the expected
probabilities and dispersion parameters of the successful attempts
is exactly the same as in Lora and Singer (2008) and it is not
shown here; we present only the estimates and standard
errors computed under the final beta-binomial model (Table 3) for comparison with
the results obtained under the
beta-binomial/gamma-Poisson model. Under this final model,
estimates of the expected probabilities of successful attempts
[$E(\pi_{gh})=\mu(\textbf z_{\mu gh})$] and dispersion parameters
$\theta(\textbf z_{\theta gh})$ (that govern the variability of
the probabilities of successful attempts), along with their
standard errors, are presented in Table 4.

\begin{table}[h]
\caption{Parameter estimates and standard errors under the final
beta-binomial model} \small \centering \vspace{0.3cm}
\begin{tabular}{llcc} \hline
             &              &            & Standard \\
   Parameter &  Related to  &  Estimate  & error \\\hline
  $\beta_{\mu0}$ & Normal group, preferred hand,  & 1.86 & 0.15 \\
  & baseline session and active sequence & & \\
  $\beta_{\mu2}$ & Effect of advanced stage & -1.35 & 0.25 \\
  $\beta_{\mu F}$ & Effect of final session & 1.38 & 0.30 \\
  $\beta_{\mu (F*C)}$ & Effect of final session and control sequence & -1.79 &
  0.30  \\ \hline
  $\beta_{\theta0}$ & Normal group, preferred hand, & -1.07 & 0.27 \\
  & baseline session and active sequence &\\
  $\beta_{\theta1}$ & Effect of initial stage & -2.98 & 1.05 \\
  $\beta_{\theta2}$ & Effect of advanced stage & 1.31 & 0.37 \\
  $\beta_{\theta (1*F)}$ & Effect of disease in initial stage and final session & 1.66 & 0.82 \\
  $\beta_{\theta (1*N)}$ & Effect of initial stage and non-preferred hand & 2.78 & 0.91 \\
  $\beta_{\theta (F*N)}$ & Effect of final session and non-preferred hand & -1.49 & 0.44 \\ \hline
\end{tabular}
\end{table}

\begin{table}[h]
\caption{Estimates of expected probabilities of successful
attempts, dispersion parameters and standard errors under the
final beta-binomial model} \vspace{0.3cm} \small \centering
\begin{tabular}{cccccc} \hline
  Disease            & Evaluation   & Intervention  &            & Expected & Standard\\
  stage              & session      & hand          & Sequence   & value    & error \\\hline
  \multicolumn {6} {c} {Expected probabilities of successful attempts} \\ \hline
  Normal or initial  & Baseline     & Either        & Either     & 0.87     & 0.02 \\
  Normal or initial  & Final        & Either        & Control    & 0.81     & 0.03 \\
  Normal or initial  & Final        & Either        & Active     & 0.96     & 0.01 \\
  Advanced           & Baseline     & Either        & Either     & 0.62     & 0.06 \\
  Advanced           & Final        & Either        & Control    & 0.52     & 0.06 \\
  Advanced           & Final        & Either        & Active     & 0.87     & 0.04 \\ \hline
  \multicolumn {6} {c} {Dispersion parameters} \\ \hline
  Normal             & Baseline     & Either        & Either     & 0.34     & 0.09 \\
  Normal             & Final        & Preferred     & Either     & 0.34     & 0.09 \\
  Normal             & Final        & Non-preferred & Either     & 0.08     & 0.03 \\
  Initial            & Baseline     & Preferred     & Either     & 0.02     & 0.02 \\
  Initial            & Baseline     & Non-preferred & Either     & 0.28     & 0.14 \\
  Initial            & Final        & Preferred     & Either     & 0.09     & 0.06 \\
  Initial            & Final        & Non-preferred & Either     & 0.33     & 0.19 \\
  Advanced           & Baseline     & Either        & Either     & 1.27     & 0.37 \\
  Advanced           & Final        & Preferred     & Either     & 1.27     & 0.37 \\
  Advanced           & Final        & Non-preferred & Either     & 0.29     & 0.13 \\\hline
\end{tabular}
\end{table}

The results suggest no evidence of difference between the
expected probabilities of successful attempts for patients using
preferred or non-preferred hand ($\beta_{\mu N}=0$), neither for
active nor for control sequences in the baseline session
($\beta_{\mu C}=0$). Patients in the normal group or with the
disease in initial stage have similar expected probabilities of
successful attempts ($\beta_{\mu 1}=0$), but those with the
disease in an advanced stage have smaller expected probabilities
of successful attempts ($\beta_{\mu 2}<0$). Moreover, an
intervention effect is detected since the expected probabilities
of successful attempts in the final session are greater than those
for the baseline session ($\beta_{\mu F}>0$). These values are
smaller for the control sequence than for the active sequence
($\beta_{\mu F}+\beta_{\mu(F*C)}<0$) suggesting that training is
effective with respect to ability.

We may also infer that there is no difference between the expected
dispersion parameter for subjects performing the active and control sequences
($\beta_{\theta C}=0$). For the normal subjects, the expected
dispersion parameters are the same ($\beta_{\theta C}$,
$\beta_{\theta N}$, $\beta_{\theta F}$=0), except in the final
evaluation using the non-preferred hand, for which the expected value
is smaller than the others ($\beta_{\theta(F*N)}<0$). For patients
in initial stage of the disease, the expected dispersion
parameters are smaller than for those in the normal group
($\beta_{\theta1}<0$); however, they change for each combination of
session and intervention hand ($\beta_{\theta(1*F)},$
$\beta_{\theta(1*N)},$ $\beta_{\theta(F*N)}\neq0$). Finally, for
patients in the advanced stage of the disease, the expected
dispersion parameter is larger than for those in the normal group
($\beta_{\theta 2}>0$), but this changes for the final session
when the non-preferred hand is used ($\beta_{\theta (F*N)}\neq0$).

\subsection{Modelling the expected number of attempts}

The initial model parameter vector, with all main effects and
first order interactions is $\boldsymbol \beta = (\boldsymbol
\beta_{\lambda}, \boldsymbol \beta_{\alpha}, \boldsymbol
\beta_{\delta})$ where
\begin{eqnarray} \nonumber
&\boldsymbol \beta_{\lambda} =&(\beta_{\lambda 0},\beta_{\lambda
1},\beta_{\lambda 2},\beta_{\lambda N},\beta_{\lambda
F},\beta_{\lambda C},
\\ \nonumber
&&\beta_{\lambda(1*F)},\beta_{\lambda(1*N)},\beta_{\lambda(1*C)},\beta_{\lambda(2*F)},\beta_{\lambda(2*N)},\beta_{\lambda(2*C)},
\beta_{\lambda(F*N)},\beta_{\lambda(F*C)},\beta_{\lambda(N*C)})\\
\nonumber &\boldsymbol \beta_{m} =&(\beta_{m 0},\beta_{m
1},\beta_{m 2},\beta_{m N},\beta_{m(1*N)},\beta_{m(2*N)})
\end{eqnarray}
with $m=\alpha, \delta$.
We may interpret $\beta_{\lambda 0}$ as the
logarithm of $\lambda$ for normal individuals, using the preferred
hand, performing the active sequence at the final evaluation;
$\beta_{\lambda N}$ corresponds to the variation in the logarithm
of $\lambda$ due to the effect of the non-preferred hand
compared to the preferred one; $\beta_{\lambda (1*N)}$
corresponds to an additional variation in the logarithm of $\lambda$ due to
the interaction between the initial stage of the
disease ($1$) and the use of the non-preferred hand ($N$). The
elements of the vector $\boldsymbol \beta_{\lambda}$ related to different evaluation sessions (represented by $F$ and $C$)  allow for
different number of attempts in these different evaluation
sessions. On the other hand, $\alpha(\textbf z_{\alpha g})$ and
$\delta(\textbf z_{\delta g})$ do not vary in different evaluation
sessions; therefore the vectors $\boldsymbol \beta_{\alpha}$ and
$\boldsymbol \beta_{\delta}$ do not have elements to distinguish
between sessions, but have elements to compare subgroups.

As noticed in Lora and Singer (2008) for the
beta-binomial model, the iterative process was very sensitive to
initial values, specially for the interactions. To overcome this
difficulty, we started with a simpler model containing only the
main effects and used the resulting estimates as initial values
for fitting other models, obtained by including the interactions
one by one. The estimates of the interaction parameters obtained
in this preliminary process were used as the initial values in our
modelling strategy.

The non-significant interactions were identified and their
simultaneous elimination from the initial model was supported
($p=0.211$) via a test of the hypothesis
\begin{eqnarray} \nonumber
&H_{0}:&\beta_{\lambda(1*F)},\beta_{\lambda(1*N)},\beta_{\lambda(1*C)},\beta_{\lambda(2*F)},\beta_{\lambda(2*N)},\beta_{\lambda(2*C)},\beta_{\lambda(F*N)},\beta_{\lambda(N*C)},\\
\nonumber
&&\beta_{\alpha(1*N)},\beta_{\alpha(2*N)},\beta_{\delta(1*N)},\beta_{\delta(2*N)}=0
\end{eqnarray}
Under the resulting reduced model, the non-significant main
effects were identified; their simultaneous elimination was
corroborated ($p=0.493$) via a test of the hypothesis
\begin{eqnarray} \nonumber
&H_{0}:&\beta_{\lambda N},\beta_{\lambda C},\beta_{\alpha
1},\beta_{\alpha 2},\beta_{\alpha N},\beta_{\delta
1},\beta_{\delta 2},\beta_{\delta N}=0.
\end{eqnarray}
We considered other hypotheses where some of these parameters are
equal to zero and they were all rejected ($p<0.150$). Goodness of
fit of the resulting reduced model was confirmed by a likelihood
ratio test in which it was compared to the initial model
($p=0.289$).

For this final model, the corresponding parameter estimates along
with their standard errors are presented in Table 5. Based on
this, we estimated expected values for $\lambda(\textbf z_{\lambda
gh})$; the results are presented in Table 6. Additionally, since
only the parameters $\beta_{\alpha 0}$ and $\beta_{\delta 0}$ were
included at the final model, we have $\alpha(\textbf z_{\alpha
g})=3.67$, with standard error of $0.18$, and $\delta(\textbf
z_{\delta g})=0.27$, with standard error of 0.07, for all disease
stages and both hands. The non-zero estimate of $\delta$ suggests
that the total attempts are overdispersed and that the
correlations among the counts across the different instants of
evaluation are non-null.

\begin{table}[h]
\caption{Parameter estimates and standard errors for the final
gamma-Poisson model} \small \centering
\begin{tabular}{llcc} \hline
  Parameter & Related to & Estimate & Standard error \\ \hline
  $\beta_{\lambda 0}$ & Normal group, preferred hand,  & 1.68 & 0.03 \\
  & initial evaluation and active sequence & \\
  $\beta_{\lambda 1}$ & Effect of initial stage & -0.38 & 0.05 \\
  $\beta_{\lambda 2}$ & Effect of advanced stage & -0.71 & 0.05 \\
  $\beta_{\lambda F}$ & Effect of final evaluation & 0.52 & 0.04 \\
  $\beta_{\lambda (F*C)}$ & Effect of final evaluation and control sequence & -0.22 & 0.05  \\ \hline
  $\beta_{\alpha 0}$ & Normal group, preferred hand & 1.30 & 0.05 \\ \hline
  $\beta_{\delta 0}$ & Normal group, preferred hand & -1.32 & 0.25 \\ \hline
\end{tabular}
\end{table}

\begin{table}[h]
\caption{Estimates of expected values of $\lambda(\textbf
z_{\lambda gh})$} \small \centering
\begin{tabular}{cccccc} \hline
  Disease   & Evaluation& Intervention & Sequence   & Expected & Standard \\
  stage     & session   & hand         &            & value    & error \\\hline
  Normal    & Baseline  & Either       & Either     & 5.4      & 0.2 \\
  Normal    & Final     & Either       & Control    & 7.2      & 0.3 \\
  Normal    & Final     & Either       & Active     & 9.0      & 0.4 \\ \hline
  Initial   & Baseline  & Either       & Either     & 3.7      & 0.2 \\
  Initial   & Final     & Either       & Control    & 5.0      & 0.4 \\
  Initial   & Final     & Either       & Active     & 6.2      & 0.3 \\ \hline
  Advanced  & Baseline  & Either       & Either     & 2.3      & 0.1 \\
  Advanced  & Final     & Either       & Control    & 3.6      & 0.2 \\
  Advanced  & Final     & Either       & Active     & 4.4      & 0.3 \\ \hline
\end{tabular}
\end{table}

We may conclude that individuals in the initial stage of the
disease have smaller expected number of attempts than normal ones,
and for individuals in the advanced stage this value is even
smaller ($\beta_{\lambda 2}<\beta_{\lambda 1}<0$ and
$\beta_{\alpha 1}=\beta_{\alpha 2}=0$). There is no evidence of
difference between the expected number of attempts for
participants using preferred or non-preferred hands
($\beta_{\lambda N}=0$ and $\beta_{\alpha N}=0$), neither for
active nor for control sequences in the baseline session
($\beta_{\lambda C}=0$). The results suggest that the training is
also effective with respect to agility, since the expected number
of attempts under the final evaluation is bigger than at the
initial one ($\beta_{\lambda F}>0$). Moreover, for the control
sequence, the expected number of attempts is larger at the final
evaluation compared with the initial one ($\beta_{\lambda
F}+\beta_{\lambda (F*C)}>0$); however, considering only the final
evaluation, the expected number of attempts is larger for the
active sequences than for the control ones ($\beta_{\lambda
(F*C)}<0$).

Table 7 contains estimates of the expected successful and total
attempts along with the respective standard errors. In Table 8 we
present estimates (with respective standard errors) of the
elements of the covariance matrix for normal subjects using the
preferred hand. Covariance patterns for the other subgroups are
similar and are not included.

\begin{table}[h]
\caption{Estimates and standard errors (within parentheses) for the
expected number of successful and total attempts under the final
beta-binomial/gamma-Poisson model} \small \centering
\begin{tabular}{cccccc} \hline
  Disease  & Evaluation & Intervention & Sequence  & Successful  & Total \\
  stage    & session    & hand         &           & attempts    & attempts   \\\hline
  Normal   & Baseline   & Either       & Either    & 17.2 (1.0)  & 19.8 (1.1) \\
  Normal   & Final      & Either       & Control   & 21.4 (0.8)  & 26.4 (0.1) \\
  Normal   & Final      & Either       & Active    & 31.7 (1.8)  & 33.0 (1.8) \\ \hline
  Initial  & Baseline   & Either       & Either    & 11.8 (0.7)  & 13.6 (0.8) \\
  Initial  & Final      & Either       & Control   & 14.9 (1.1)  & 18.4 (1.2) \\
  Initial  & Final      & Either       & Active    & 21.9 (1.4)  & 22.8 (1.4) \\\hline
  Advanced & Baseline   & Either       & Either    &  5.2 (0.6)  & 8.4 (0.6) \\
  Advanced & Final      & Either       & Control   &  6.9 (0.9)  & 13.2 (0.9) \\
  Advanced & Final      & Either       & Active    & 14.0 (1.2)  & 16.1 (1.1) \\\hline
\end{tabular}
\end{table}

\begin{table}[h]
\caption{ Estimates and standard errors (within parentheses) for
the expected covariance matrix for normal subjects using the
preferred hand} \small \centering
\begin{tabular}{cccccccccccc} \hline
   & & & \multicolumn{4}{c}{Baseline session} &\quad \qquad& \multicolumn{4}{c}{Final session} \\
   & & & \multicolumn{2}{c}{Active seq.}& \multicolumn{2}{c}{Control
   seq.}&&   \multicolumn{2}{c}{Active seq.}& \multicolumn{2}{c}{Control seq.}  \\
   & & & Suc. & Att. & Suc. & Att. & &Suc. & Att. & Suc. & Att.  \\ \hline
   Baseline & Active & Suc. & 51.2 & & & &&  &  & &  \\
   session &seq.&& (7.2)& & & &&  &  & &  \\
   & & Att. & 42.2 & 48.5 & & &&  &  & &  \\
   & &  & (11.4) & (13.0)& & &&  &  & &  \\
   & Control & Suc. & 21.5  & 0 & 51.2& &&  &  & &  \\
   & seq. & & (5.8) &  & (7.2)& &&  &  & &  \\
   &  & Att. & 0 & 28.7 & 42.4 & 48.5 &&  &  & &  \\
   &  &  & & (7.8) & (11.4) & (13.0) &&  &  & &  \\
   &&&&&&&&&& \\
   Final & Active & Suc. & 40.3 & 0 & 40.3 & 0&& 118.9 &  & &  \\
   session & seq. &  & (10.8) &  & (10.8) & && (22.2) &  & &  \\
   & & Att. & 0 & 48.4  & 0 & 48.4& & 110.5 &  115.1& &  \\
   & & & & (13.0) &  &  (13.0)& & (30.0)&   (31.2)& &  \\
   & Control & Suc. & 27.3 & 0 & 27.3 & 0 & &52.3& 0 & 87.4  &   \\
   & seq. &  &  (7.3) &  &  (7.3) &  & & (13.8)&  & (12.9) &   \\
   & & Att. & 0 & 39.0 & 0& 39.0  & & 0& 65.8  & 64.9& 80.1  \\
   & & &  & (10.5) & & (10.5) & & & (17.7) & (17.8)& (21.8) \\ \hline
\end{tabular}
\begin{center}
Codes: Suc.$=$Successes, Att.$=$Attempts and seq.$=$sequence
\end{center}
\end{table}

\section{Discussion}

The proposed beta-binomial/gamma-Poisson model is more general
than the multivariate beta-binomial/Poisson model considered in
Lora and Singer (2008) because it allows for different covariances
between the number of attempts in different evaluation sessions
and considers a possible overdispersion of the total attempts.
Moreover, the gamma-Poisson component of the model is
computationally much easier to use for comparisons among the
numbers of attempts in different evaluation sessions.

While in the multivariate beta-binomial/Poisson model, the multivariate Poisson
component requires a different set of parameters for each
evaluation session, in the beta-binomial/gamma-Poisson model, the
gamma-Poisson component includes a single set of parameters for
all evaluation sessions. To compare the expected number of
attempts under different conditions using the former, it is
necessary rewrite the model and to derive ad hoc estimating
equations while under the latter, it suffices to eliminate the
corresponding regression parameter and to obtain new parameter
estimates using the same estimating equations. For the analyzed
data, for example, the comparison between the control and active
sequence during the baseline evaluation using the
beta-binomial/gamma-Poisson model is done by testing if the
parameter $\beta_{\lambda C}$ is null. On the other hand, under the
multivariate beta-binomial/Poisson approach, the total
number of trials is modelled with a specific vector of parameters for each instant
of observation; for the data in the example, they are: baseline
evaluation performing active sequence, baseline evaluation
performing the control sequence, final evaluation performing the
active sequence and final evaluation performing the control
sequence. To compare the control and active sequences during
the baseline session we should rewrite the model using only three
parameters: baseline evaluation (the same for active and control sequences), final
evaluation performing active sequence and final evaluation
performing control sequence.

The average of the absolute differences between the sample means
of the number of successful and total attempts and the respective
expected values under this final model (Table 7) is 1.7. The same
average based on the multivariate beta-binomial/Poisson model is
0.9. Furthermore, the average of the absolute differences between
the observed and estimated covariances using the multivariate
beta-binomial/Poisson model is 21.5 while it is 19.1 if we use the
beta-binomial/gamma-Poisson. These differences are attributable to
the more flexible covariance structure induced by the latter,
i.e., allowing for different covariances between the repeated
number of trials.

The values of the AIC ( = 1888.0) and the BIC ( = 1919.1) for the
beta-binomial/gamma-Poisson model compared to the corresponding
values (AIC = 1935.6 and BIC = 1974.0) for the multivariate
beta-binomial/Poisson also suggest a better fit of the former.

Although the results are quite similar, with the exception of the
values for patients in the advanced stage of the disease, the
beta-binomial/gamma-Poisson one is preferable to the multivariate
beta-binomial/Poisson, both because of the modelling flexibility
and the computational advantages mentioned before.

As an extension for the beta-binomial/gamma-Poisson model, we could
incorporate a parameter to relate the probabilities of success to
the total attempts, as in Zhu et al. (2004).
Another possible
extension would be to consider the case where attempts could be done
correctly, satisfactorily or incorrectly; in this case, we could
generalize the model by considering
Dirichlet-multinomial/gamma-Poisson distribution models. These extensions
are currently under investigation.

\section*{Appendix}

\noindent \textbf{First and second derivatives for the
gamma-Poisson model}

$$\frac{\partial L(\boldsymbol \beta_{\lambda}, \boldsymbol \beta_{\alpha},
\boldsymbol \beta_{\delta})}{\partial \boldsymbol
\beta_{\lambda}}= \textbf Z_{\lambda}' \textbf L \left [ \textbf
L^{-1} \textbf n - (\textbf I_{p} \otimes \textbf B^{-1}) (\textbf
1_{p} \otimes \textbf a) \right ] ,\:
$$
$$\frac{\partial L(\boldsymbol \beta_{\lambda}, \boldsymbol \beta_{\alpha},
\boldsymbol \beta_{\delta})}{\partial\boldsymbol
\beta_{\alpha}}=\textbf Z_{\alpha}' \textbf M [\textbf c - \textbf
D^{-1} \textbf {log} (\textbf b)] \:\: {\rm and}$$
$$
\frac{\partial L(\boldsymbol \beta_{\lambda}, \boldsymbol
\beta_{\alpha}, \boldsymbol \beta_{\delta})}{\partial\boldsymbol
\beta_{\delta}}=\textbf Z_{\delta}' [\textbf {De} + \textbf D^{-1}
\textbf M \textbf {log} (\textbf b) - \textbf B^{-1} \textbf L_{s}
\textbf a]
$$
$$\frac{\partial^{2}L(\boldsymbol \beta_{\lambda}, \boldsymbol \beta_{\alpha},
\boldsymbol \beta_{\delta})}{\partial \boldsymbol \beta_{\lambda}
\partial \boldsymbol \beta_{\lambda}'}= \textbf Z_{\lambda}'
\textbf L [\textbf I_{p} \otimes (\textbf {AB}^{-1})] \left \{
\textbf L [\textbf I_{p} \otimes (\textbf {DB}^{-1})] - \textbf
I_{Mp} \right \} \textbf Z_{\lambda},$$
$$\frac{\partial^{2}L(\boldsymbol \beta_{\lambda}, \boldsymbol \beta_{\alpha},
\boldsymbol \beta_{\delta})}{\partial \boldsymbol \beta_{\lambda}
\partial \boldsymbol \beta_{\alpha}'}= -\textbf Z_{\lambda}' \textbf L
[\textbf I_{p} \otimes (\textbf {MB}^{-1})]  (\textbf 1_{p}
\otimes \textbf Z_{\alpha}),$$
$$\frac{\partial^{2}L(\boldsymbol \beta_{\lambda}, \boldsymbol \beta_{\alpha},
\boldsymbol \beta_{\delta})}{\partial \boldsymbol \beta_{\lambda}
\partial \boldsymbol \beta_{\delta}'}= -\textbf Z_{\lambda}'
\textbf L [\textbf I_{p} \otimes (\textbf {DB}^{-2})] \left [
\textbf I_{p} \otimes (\textbf N_{s}-\textbf M \textbf L_{s})
\right ] (\textbf 1_{p} \otimes \textbf Z_{\delta}) ,$$
$$\frac{\partial^{2}L(\boldsymbol \beta_{\lambda}, \boldsymbol \beta_{\alpha},
\boldsymbol \beta_{\delta})}{\partial \boldsymbol \beta_{\alpha}
\partial \boldsymbol \beta_{\alpha}'}= \textbf Z_{\alpha}' \textbf
M \left [ \textbf C - \textbf D^{-1} \textbf{log}(\textbf B) -
\textbf {MF} \right ] \textbf Z_{\alpha},$$
$$\frac{\partial^{2}L(\boldsymbol \beta_{\lambda}, \boldsymbol \beta_{\alpha},
\boldsymbol \beta_{\delta})}{\partial \boldsymbol \beta_{\alpha}
\partial \boldsymbol \beta_{\delta}'}= \textbf Z_{\alpha}' \textbf
{MD} \left [- \textbf J - \textbf D^{-1} \textbf B^{-1} \textbf
L_{s} + \textbf D^{-2} \textbf{log}(\textbf B) \right ]\textbf
Z_{\delta} \:\: {\rm and}$$
$$\frac{\partial^{2}L(\boldsymbol \beta_{\lambda}, \boldsymbol \beta_{\alpha},
\boldsymbol \beta_{\delta})}{\partial \boldsymbol \beta_{\delta}
\partial \boldsymbol \beta_{\delta}'}= \textbf Z_{\delta}' \textbf
D \left \{ \textbf E - \textbf {DQ} + \textbf M \textbf D^{-1}
\textbf B^{-1} \textbf L_{s} - \textbf D^{-2} \textbf M
\textbf{log}(\textbf B) - \textbf B^{-2} \textbf L_{s} \left [
\textbf N_{s} - \textbf M \textbf L_{s} \right ] \right \} \textbf
Z_{\delta}$$
with $L(\boldsymbol \beta_{\lambda}, \boldsymbol \beta_{\alpha},
\boldsymbol \beta_{\delta})$ presented in (24) and
\begin{eqnarray} \nonumber
&&\textbf a=(a_{1},...,a_{g},...,a_{M})',\:\: a_{g}=\delta
(\textbf z_{\delta g}) \left [\sum_{h=1}^{p} n_{gh} \right] +
\alpha (\textbf z _{\alpha g})
\\ \nonumber
\nonumber && \textbf A=diag\{a_{g}\}
\\ \nonumber
&&\textbf B=diag\{b_{g}\},\:\:b_{g}=\delta (\textbf z_{\delta g})
\left [\sum_{h=1}^{p} \lambda (\textbf z_{\lambda gh}) \right]+1
\\\nonumber && \textbf {log}(\textbf
b)=(log(b_{1}),...,log(b_{g}),...,log(b_{M}))'
\\ \nonumber
&& \textbf {log}(\textbf B)= diag\{log(b_{g})\}
\\
\nonumber && \textbf c=(c_{1},...,c_{g},...,c_{M})',\:\:
c_{g}=\sum_{u=0}^{\Sigma_{h=1}^{p}n_{gh}-1} \frac{1}{\alpha
(\textbf z_{\alpha g})+u\delta (\textbf z_{\delta g})
},\\\nonumber &&\textbf C = diag\{c_{g}\}
\end{eqnarray}

\begin{eqnarray} \nonumber
&& \textbf e=(e_{1},...,e_{g},...,e_{M})',\:\:
e_{g}=\sum_{u=0}^{\Sigma_{h=1}^{p}n_{gh}-1} \frac{u}{\alpha
(\textbf z_{\alpha g})+u\delta (\textbf z_{\delta g}) }
\\ \nonumber
&&\textbf E = diag\{e_{g}\},
\\ \nonumber
&&\textbf F =diag\{f_{g}\},\:\:f_{g}=
\sum_{u=0}^{\Sigma_{h=1}^{p}n_{gh}-1} \frac{1}{[\alpha (\textbf
z_{\alpha g})+u\delta (\textbf z_{\delta g})]^{2}}
\\\nonumber
&& \textbf J = diag\{j_{g}\},\:\:j_{g}=
\sum_{u=0}^{\Sigma_{h=1}^{p}n_{gh}-1} \frac{u}{[\alpha (\textbf
z_{\alpha g})+u\delta (\textbf z_{\delta g})]^{2}}
\\\nonumber
&& \textbf Q = diag\{q_{g}\},\:\:q_{g}=
\sum_{u=0}^{\Sigma_{h=1}^{p}n_{gh}-1} \left [ \frac{u}{\alpha
(\textbf z_{\alpha g})+u\delta (\textbf z_{\delta g})} \right
]^{2}
\\ \nonumber
&& \textbf n=(n_{11},...,n_{gh},...,n_{Mp})', \\ \nonumber &&
\textbf N_{s}=diag\left\{\sum_{h=1}^{p} n_{gh}\right\}
\\\nonumber
&& \textbf L=diag\{\lambda (\textbf z_{\lambda gh})\}
\\\nonumber && \textbf
L_{s}=diag\left\{\sum_{h=1}^{p}\lambda (\textbf
 z_{\lambda gh})\right\}
 \\\nonumber && \textbf M=diag\{\alpha
(\textbf z_{\alpha g})\}
\\\nonumber && \textbf D=diag\{\delta (\textbf z_{\delta g})\}
\\\nonumber && \textbf Z_{\lambda}=(\textbf z_{\lambda
11}',...,\textbf z_{\lambda gh}',...,\textbf z_{\lambda Mp}')'
\\\nonumber && \textbf Z_{\alpha}=(\textbf z_{\alpha 1}',...,\textbf
z_{\alpha g}',...,\textbf z_{\alpha M}')'
\\\nonumber && \textbf Z_{\delta}=(\textbf z_{\delta 1}',...,\textbf
z_{\delta g}',...,\textbf z_{\delta M}')'
\end{eqnarray}

\section*{Acknowledgements}

We are grateful to Conselho Nacional de Desenvolvimento Científico
e Tecnológico (CNPq) and Fundação de Amparo à Pesquisa do Estado
de São Paulo (FAPESP), Brazil, for partial financial support. We
are also grateful to Maria Elisa Pimentel Piemonte for providing
the data.

\end{document}